\documentclass[bmf, amsmath,amssymb, preprint]{revtex4-1}
\usepackage{fancyhdr}
\usepackage{graphicx}
\usepackage{lipsum} 
\usepackage{amsmath}
\usepackage{xcolor}
\usepackage{dcolumn}
\usepackage{bm}
\usepackage{hyperref}
\usepackage[mathlines]{lineno}
\usepackage{mathtools}
\usepackage{cuted}
\usepackage{float}
\usepackage{afterpage}

\begin{document}
\title{Roughness-induced critical phenomenon analogy for turbulent friction factor  explained by a co-spectral budget model}
\author{Shuolin Li}
\email{shuolin.li@duke.edu}
\affiliation{Department of Civil and Environmental Engineering and Nicholas School of the Environment, Duke University, Durham, NC, USA}
\author{Gabriel Katul}
\email{gaby@duke.edu}
\affiliation{Department of Civil and Environmental Engineering and Nicholas School of the Environment, Duke University, Durham, NC, USA}

\begin{abstract}
Drawing on an analogy to critical phenomena, it was shown that the Nikuradse turbulent friction factor ($f_t$) measurements in pipes of radius $R$ and wall roughness $r$ can be collapsed onto a one-dimensional curve expressed as a conveyance law $f_t Re^{1/4}=g_o(\chi)$, where $Re$ is a bulk Reynolds number, $\chi =Re^{3/4}\left({r}/{R}\right)$. The implicit function $g_o(.)$ was conjectured based on matching two asymptotic limits of $f_t$. However, the connection between $g_o(.)$ and the phenomenon it proclaims to represent - turbulent eddies - remains lacking.  Using models for the wall-normal velocity spectrum and return-to-isotropy for pressure-strain effects to close a co-spectral density budget, a derivation of $g_o(.)$ is offered. The proposed method explicitly derives the solution of the conveyance law and provides a physical interpretation of $\chi$ as a dimensionless length scale reflecting the competition between viscous sublayer thickness and characteristic height of roughness elements. The application of the proposed method to other published measurements spanning roughness and Reynolds numbers beyond the original Nikuradse range is further discussed.
\end{abstract}
\maketitle

\section{Introduction}
\begin{figure}
\centerline{\includegraphics[angle=0,width=0.99\linewidth]{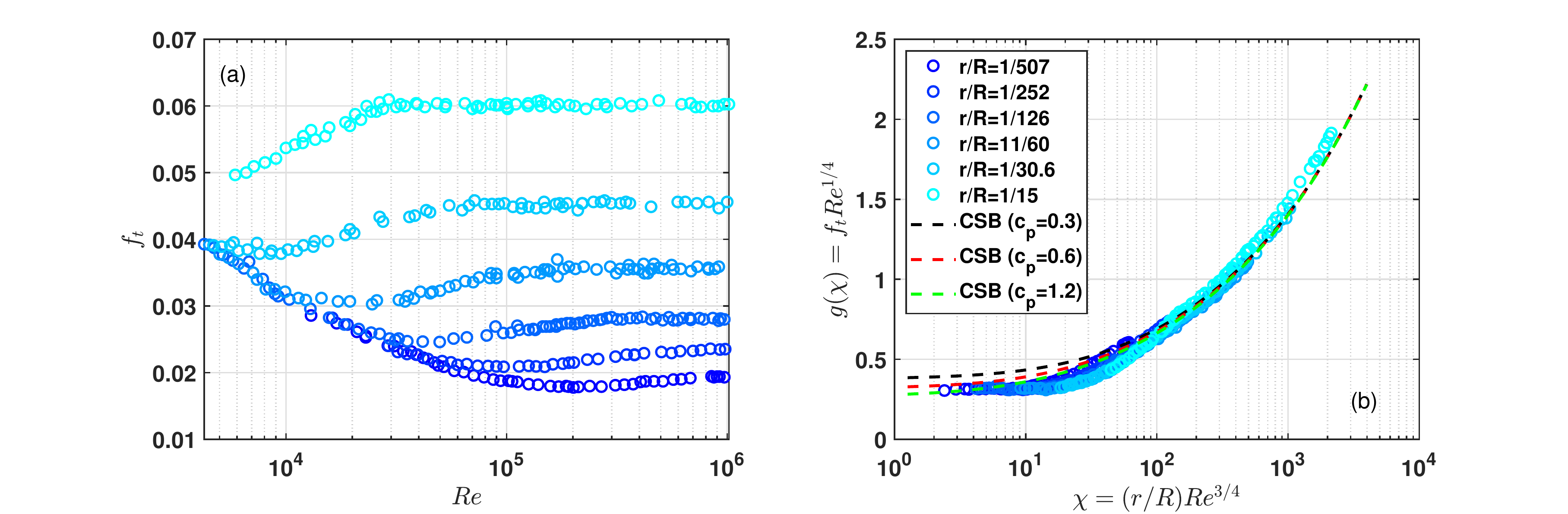}}
\caption{The original Nikuradse diagram and its NG06 representation with $f_t Re^{1/4}$ expressed as an unknown function of $\chi$. The coefficient $c_p=0.61$ can be theoretically determined when combining Blausius and Strickler scaling \cite{gioia2001scaling}.}
\label{fig:univ1}
\end{figure}

A recent analogy between critical phenomena and turbulent flows was proposed to describe the turbulent friction factor $f_t $ in pipes \citep{goldenfeld2006roughness}.  The $f_t$ is is a dimensionless measure of the total frictional loss defined as 
\begin{equation}
f_{t} =\frac{g R S_b}{\frac{1}{2}U_b^2},
\label{eq:f_def}
\end{equation}
and is presumed to vary with the bulk Reynolds number ($Re=U_b 2R/\nu$) and relative roughness of the wall ($r/R$), where $g$ is the gravitational acceleration, $U_b$ is the bulk or time and area-averaged velocity, $r$ is a measure of the wall roughness often related to the statistics of the protrusions from the pipe wall, $R$ is the pipe radius, $\nu$ is kinetic viscosity, and $S_b$ is the friction slope that can be related to the driving force - the mean pressure gradient \citep{clark2011transport}. Using measured $f_t$, the weighty experiments by Nikuradse \citep{Nikuradse} on regular roughness elements identified two limiting flow regimes - hydrodynamically smooth and fully-rough based on competing mechanisms between $r/R$ and a length scale measuring the thickness of the viscous sublayer $L_v= 5\eta$, where $\eta$ is the Kolmogorov micro-scale. In the hydrodynamically smooth case (i.e. $r/L_v\ll1$), $f_t=A_bRe^{-1/4}$ where $A_s=0.316$ (labelled as the Blasius scaling) whereas in the fully-rough regime ($r/L_v\gg1$), $f_t=A_s (r/R)^{1/3}$ where $A_s=0.14$ (labelled as the Strickler scaling).  Exploiting an analogy developed to infer thermodynamic properties of ferromagnets near critical temperatures, the Nikuradse's $f_t$ data was shown to collapse (albeit imperfectly) onto a single curve labeled here as NG06 \citep{goldenfeld2006roughness}. In the derivation of NG06, two limiting regimes occurring for $r/R \rightarrow 0$ (analogous to an external magnetic field control) and $Re^{-1} \rightarrow 0$ (analogous to inverse temperature near its critical state), respectively, have been exploited.  The NG06 $f_t$ was shown to be mathematically described by \citep{goldenfeld2006roughness}
\begin{equation} 
\label{eq:ng06}    
f_t=Re^{-1/4}g(\chi), ~~\mathrm{with}~ g_o(\chi) = \left\{
\begin{array}{ll}
\mathrm{const.}, & \chi \rightarrow 0 \\[2pt]
\chi^{1/3},        & \chi\rightarrow \infty 
\end{array} \right.,
\end{equation}
where $\chi =Re^{3/4}\left({r}/{R}\right)$ and $g_o(.)$ is an implicit function satisfying two asymptotic properties: when $\chi\rightarrow0$, equation \ref{eq:ng06} yields $f_t\sim Re^{-1/4}$ (the Blasius scaling) whereas at sufficiently large $Re$, $g_o(\chi)$ becomes $Re^{1/4}(r/R)^{1/3}$ resulting in $f_t\sim (r/R)^{1/3}$ (the Strickler scaling).  The outcome of equation \ref{eq:ng06} is a monotonic curve along which all the Nikuradse data collapse as shown in Figure \ref{fig:univ1}

The collapse of all Nikuradse data when representing $f_t Re^{1/4}$ as a function of $\chi$ may indicate the existence of a critical phenomenon in the turbulent friction factor \citep{goldenfeld2006roughness}. The NG06 stimulated other theories and a combination of variables derived from hydrodynamic stability analysis for laminar flow \citep{tao2009critical, li2016united}. Other approaches to deriving $f_t$ and refinements to NG06 exploited the so-called spectral link in turbulent flows \citep{gioia2001scaling,gioia2006turbulent,MehrafarinPourtolami08,guttenberg2009friction,calzetta2009friction,goldenfeld2017turbulence,anbarlooei2020new} summarized by
\begin{equation}
f_{t} \propto \sqrt{\int_{1/l_o}^{\infty}{E_{TKE}(k)dk}},
\label{eq:f_Etke0}
\end{equation}
where $k$ is wavenumber or inverse eddy size, $l_o=r+L_v$, and $E_{TKE}(k)$ is the spectrum of the turbulent kinetic energy (TKE). This relation was tested using two-dimensional soap film experiments where $E_{TKE}(k)$ was manipulated to scale as $k^{-5/3}$ or $k^{-3}$ depending on whether the inverse energy cascade or forward enstrophy (or integral of vorticity) cascade applies \citep{guttenberg2009friction,tran2010macroscopic,kellay2012testing}. Another corollary improvement to NG06 routed in equation \ref{eq:f_Etke0} was intermittency corrections to phenomenological models for $E_{TKE}(k)$ \citep{MehrafarinPourtolami08}.  When such intermittency corrections are accounted for in $E_{TKE}(k)$, a revised NG06 that better describes the otherwise imperfect fit was reported.  The collapse of the Nikuradse data onto a single (albeit in a restricted range of $r/R-Re$) curve is appealing because it offers a diagnostic description of the so-called transitional regime between smooth and fully rough cases \citep{gioia2001scaling} or other similarity variants on it \citep{li2016united}. That transitional flow regimes in $f_t$ exhibit rich scaling laws are now opening up new vistas to other analogies in physics and statistical mechanics \citep{goldenfeld2017turbulence} though no contact with Navier-Stokes turbulence or approximations to it has been offered to date. 

This work explains $g_o(\chi)$ and derives its generalization for steady and axially uniform turbulent pipe flow using standard turbulent theories. The theoretical tactic employs a co-spectral budget (CSB) model that makes contact with an approximated Navier-Stokes equation for the near-wall turbulent stress in spectral space \citep{katul2013co,katul2014cospectral,bonetti2017manning,coscarella2021relation}. The outcome is an analytical formulation linking an externally specified wall-normal energy spectrum to the turbulent stress (via the CSB model), and upon scale-wise integration yields an expression for $f_t$ analogous in form to equation \ref{eq:f_Etke0}. This expression includes a bridge between local variables formulated on a plane positioned at a wall distance $z_*$ that scales with $l_o$ and bulk flow variables reflecting the overall geometry and flow rate in the pipe \citep{bonetti2017manning,coscarella2021relation}.  The proposed model is shown to collapse the expanded $f_t$ data onto a single curve whose shape is explicitly derived from the CSB model with similarity constants all linked to standard constants in turbulence theories.  Other mechanisms not explicitly treated such as intermittency corrections \citep{MehrafarinPourtolami08} (or other similarity variants \citep{li2016united}), to the wall-normal velocity spectrum, non-local spectral transfer across scales in energy and stresses, non-linear return-to-isotropy representations for pressure-velocity interactions, or bottle-necks in the energy cascade can all be accommodated in this framework and their effects tracked onto an NG06 type curve but they are not explicitly considered here.

\section{Theory}
\subsection{Definitions}
\begin{figure}
\centerline{\includegraphics[angle=0,width=0.99\linewidth]{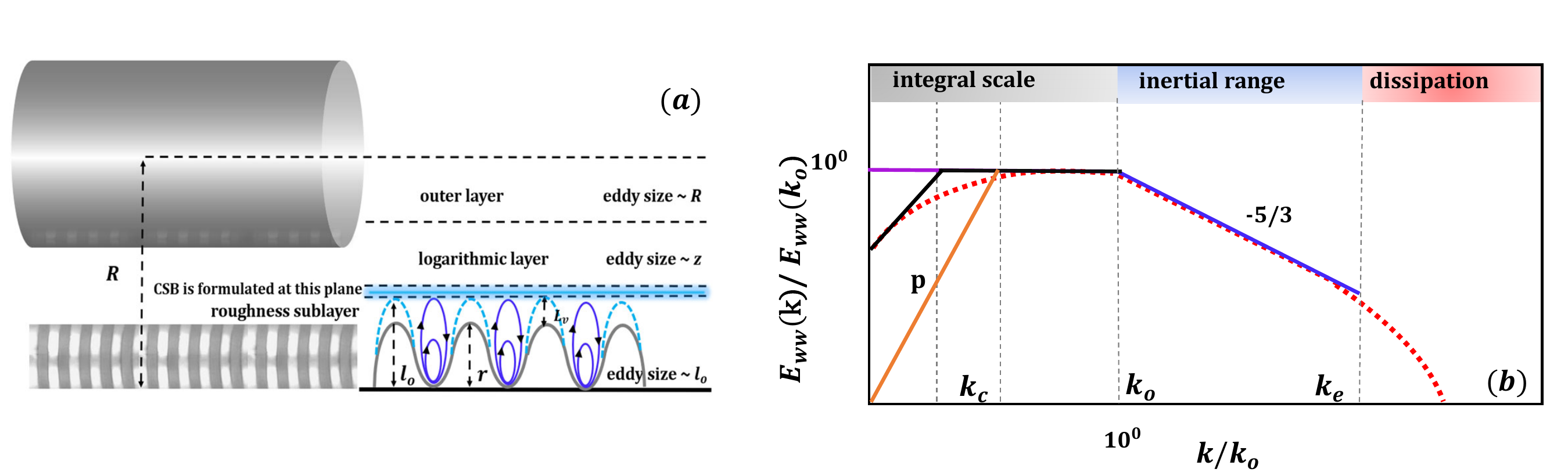}}
\caption{(a): The formulation of the CSB model at $z_*/l_o>1$ but below the log-region. (b): Schematic of the wall-normal velocity spectrum $E_{ww}(k)$ of a turbulent rough pipe as a function of wavenumber $k$. The dashed red line represents commonly observed $E_{ww}(k)$ that can be partitioned into three subranges: an energetic range (integral scale), inertial range, and viscous dissipation range. The solid lines show different models for this spectrum. Plausibility check of the assumed $E_{ww}(k)$ shape can be made using integral constrain shown in Supplementary Material.}
\label{fig:etke}
\end{figure}

The flow is assumed to be stationary and longitudinally homogeneous driven by a constant mean pressure-gradient within a pipe of radius $R$ and cross-sectional area $A_p$. The pipe wall is uniformly covered with regular roughness elements having a protrusion amplitude $r$ similar to the Nikuradse experiments (see Figure \ref{fig:etke}a). 
Defining $z=R-y$ as the normal distance to the pipe boundary, $y$ as the distance from the pipe center, $U^+=U(z)/u_{*}$ as the dimensionless mean velocity profile, $u_*=\sqrt{\tau_o/\rho_f}$ as the friction velocity, $\tau_o$ as the wall stress, $\rho_f$ as the fluid density, the bulk (i.e. time and cross-sectional area-averaged) velocity can be determined as 
\begin{equation}
\label{Ub_def}
U_b=\frac{1}{A_p} \int_0^R U(y)dA_s, 
\end{equation}
where $dA_s=2\pi y dy$. For this setup, the mean longitudinal momentum balance reduces to a balance between the mean pressure gradient and the stress gradient given by
\begin{equation}
\frac{\partial P}{\partial x}=\frac{1}{y}\frac{\partial (y \tau)}{\partial y}, 
\end{equation}
where $P$ is the mean pressure, $x$ is the longitudinal distance along the pipe length, and $\tau(y)$ is the total shear stress at radial distance $y$ from the pipe center.  Integrating with respect to $y$ yields 
\begin{equation}
\tau(y)=\left(\frac{\partial P}{\partial x}\right)\frac{y}{2}  + C_1,
\end{equation}
where $C_1$ is determined so that at $y=0$ (i.e. center of the pipe), $\tau(0)=0$ due to symmetry thereby resulting in 
\begin{equation}
\tau(y)=\left(\frac{\partial P}{\partial x}\right)\frac{y}{2}. \end{equation}
Defining 
\begin{equation}
u_*^2=\frac{\tau_o}{\rho_f}=\frac{1}{\rho_f}\frac{R}{2} \left(\frac{\partial P}{\partial x}\right)=\frac{R}{4}gS_b,
\label{eq:ghs}
\end{equation}
\noindent and decomposing $\tau(y)$ into a turbulent $\tau_{t}$ and a viscous $\tau_{m}$ contribution leads to the variation in total stress with distance from the wall as 
\begin{equation}
\tau(z)=\tau_{t}+\tau_{m}=\tau_o \left(1-\frac{z}{R}\right), 
\end{equation}
where $\tau_{m}=\rho_f\nu \Gamma(z)$ and $\Gamma(z)=dU/dz$.  

\subsection{The co-spectral budget model}
The CSB model is now formulated at a wall-normal distance $z=z_*$ below the region where the onset of a logarithmic mean velocity profile for $U^+=U(z)/u_*$ is expected (Figure \ref{fig:etke}). Hence, the effective eddy size $l_o$ impacting momentum exchange at $z_*$ need not scale with $z$ \citep{katul2014cospectral, bonetti2017manning} but with $r$ or $L_v$ ($=5\eta/2$) depending on whether the flow is rough or smooth. To accommodate rough and smooth pipe flow conditions, we define $l_o = r+L_v$ as before \citep{gioia2006turbulent,coscarella2021relation}, where $\eta=(\nu^3/\epsilon)^{1/4}$, and $\epsilon$ is a local turbulent kinetic energy dissipation rate evaluated at $z_*$. A justification for summing $r$ and $L_v$ is that resistances to momentum exchanges between a moving fluid and a stationary wall in the combined layer are additive, and resistances scale linearly with layer thicknesses. The following constraints on the choice of $z_*$ are now enforced: $z_*/l_o \ge 1$ and $\tau_t(z_*)/\tau_o \approx 1$, where $\tau_t(z_*)$ is the turbulent stress at $z_*$ labelled hereafter as $\tau_*$ for notation convenience.  Selecting $z_*$ to be sufficiently distant from the boundary also minimizes wall-blocking effects impacting $U^+$ in the buffer region \citep{MccollEA16}. For a rough pipe, the $z_*$ is still expected to be in the roughness sublayer (RSL) whereas in a turbulent smooth pipe, $z_*$ is in the upper-region of the buffer layer \citep{Raupach81, RaupachEA91,pope2001turbulent, PoggiEA02}.  

The CSB model links $\tau_*$ to eddy sizes at $z_*$ using
${\tau_*}/{\rho_f}=-\overline{{u}'{w}'} =\int_{0}^{\infty }F_{uw}(k)dk,$
and $F_{uw}(k)$ defines the co-spectrum between $u'$ and $w'$ at $k$ (or inverse eddy-size), $u'$ and $w'$ are the turbulent longitudinal and wall-normal velocity components, respectively, primed quantities are excursions from the mean state, and overline is averaging over coordinates of statistical homogeneity (usually surrogated to time averaging). The terms governing the time evolution of the co-spectral budget at $z_*$ are \citep{katul2013co,katul2014cospectral}
\begin{equation} \label{eq:csb}    
\frac{\partial F_{uw}(k)}{\partial t}=P_{uw}(k)+T_{uw}(k) +\left[\pi_{uw}(k)-D_{uw}(k)\right],
\end{equation}
where $P_{uw}(k)=\Gamma (z)E_{ww}(k)$ is the turbulent stress production term at wavenumber $k$ due to the presence of a mean velocity gradient $\Gamma$, $E_{ww}(k)$ is the energy spectrum of the wall-normal velocity component, $T_{uw}(k)$ is a scale-wise transfer of momentum and satisfies $\int_{0}^{\infty }T_{uw}(k) dk=0$, $\pi_{uw}(k)$ is a pressure-velocity de-correlation term commonly modeled using return to isotropy principles, and $D_{uw}(k)=2\nu k^2 F_{uw}(k)$ is a viscous dissipation term also responsible for de-correlating $u'$ from $w'$. Stationarity is assumed through out and closure models for $\pi_{uw}(k)$ and $T_{uw}(k)$ are needed.  For maximum simplicity and to ensure a recovery of $F_{uw}(k) \propto k^{-7/3}$ in the so-called inertial subrange (ISR), $T_{uw}(k)=0$ is assumed (and justified later on).  Adopting a linear Rotta scheme revised for isotropization of the production at any $k$, $\pi_{uw}(k)$ is closed by \citep{katul2013co,katul2014cospectral,bonetti2017manning,MccollEA16,coscarella2021relation}
\begin{equation}
\pi_{uw}(k)=-C_R\frac{1}{t_r(k)}F_{uw}(k)-C_IP_{uw}(k),
\end{equation}
where $C_R\approx 1.8$ and $C_I=3/5$ \citep{pope2001turbulent} are the Rotta and isotropization of production constants, and $t_r(k)=[k^3E_{kol}(k)]^{-1/2}$ is a local wavenumber dependent relaxation time scale \citep{onsager1949statistical,pope2001turbulent} based on a Kolmogorov spectrum $E_{kol}(k)$. Other possibilities that include non-local energy transfer can be accommodated using $t_r(k)$. For example, a non-local closure for the energy flux is the Heisenberg model \citep{heisenberg1948theory} that can be re-casted as $t_r(k)^{-1}=\sqrt{\int_0^k p^2 E_{kol}(p)dp}$ \citep{katul2012existence}. There are issues with the Heisenberg model related to the directional energy transfer and equipartition of energy that have already been identified and discussed \citep{clark2009reassessment}.  For this reason, the focus here is maintained on the simpler $t_r(k)=[k^3E_{kol}(k)]^{-1/2}$.  The $t_r(k)$ becomes unbounded as $k \rightarrow 0$ necessitating additional constraints at large scales.  One possible constraint is to set $t_r(k)=t_r(k_c)$ when $k/k_c<1$ where $k_c = 1/R$ is the smallest inverse length scale over which $E_{ww}(k)$ energy transfer occurs downscale.  The two destruction terms in the CSB model, the Rotta component of $\pi_{uw}(k)$ and viscous destruction $D_{uw}(k)$, are compared at small scales (or large $k$) using
\begin{equation}
\Phi(k)=\frac{2\nu k^2 F_{uw}(k)}{C_R F_{uw}(k)/t_r(k)}=\frac{2}{C_R} (k\eta)^{4/3},
\end{equation}
where the role of $P_{uw}(k)$ has been ignored at large $k$ for simplicity. When $k\eta\ll1$, the viscous dissipation is negligible ($\Phi(k)\approx 0$) compared with the Rotta term. However, as $k\eta>1$, the viscous term dominates and $\Phi^{-1}(k)\approx 0$. Adopting the aforementioned closure schemes, the co-spectrum at $k$ is derived as
\begin{equation} \label{eq:fuw}    
F_{uw}(k)=\frac{1-C_I}{C_R}\Gamma (z_*)\frac{ E_{ww}(k) t_r(k)}{\Phi(k)+1}.
\end{equation}
The co-spectrum must be integrated across all $k$ to yield $\tau_*$ needed in the determination of $f_t$. To evaluate $F_{uw}(k)$, the shape of $E_{ww}(k)$ is required and discussed in Figure \ref{fig:etke}. 

The $E_{ww}(k)$ in the ISR is given by the Kolmogorov spectrum $E_{kol}(k)=C_o\left[\epsilon(z_*)\right]^{2/3} k^{-5/3}$ where $C_o=(24/55)C_o'$ is the Kolmogorov constant for the wall-normal velocity component \citep{pope2001turbulent} related to the Kolmogorov constant for the turbulent kinetic energy spectrum ($C_o'=1.5$). Deviations from $E_{kol}$ at other scales are specified as follows: i) an exponential cutoff, $f_{\eta}=\exp\left[-\beta\left(\frac {1}{2}C_R\Phi\right)^{3/4}\right]\approx 1$, when $k\ll k_e$, where $\beta=2.1$, to resolve the viscous dissipation range \citep{pope2001turbulent, gioia2010spectral}; ii) two piece-wise functions for the energetic range \citep{bonetti2017manning}. The $E_{ww}(k)$ is
\begin{equation} 
\label{eq:ewweq} 
E_{ww}(k)= \left\{ \begin{array}{l l} 
                E_{kol}(k_o)k_c^{-p}k^p & \mathrm{if} ~0\leqslant k \leqslant  k_c\\
E_{kol}(k_o) & \mathrm{if} ~k_c\leqslant k \leqslant  k_o\\
E_{kol}(k)f_{\eta} & \mathrm{otherwise}
\end{array} \right.
\end{equation}
where $k_c = 1/R$, $k_o=1/l_o$, $k_e = 1/\eta$ are three characteristic wavenumbers that mark the key transitions in $E_{ww}(k)$ as related to pipe radius, characteristic eddy scale in the RSL, and the viscous length scale \citep{bonetti2017manning, katul2013co,li2019cospectral, pope2001turbulent,coscarella2021relation}, and several theories constrain $p$ to be between 2 (Saffman spectrum) and 4 (Batchelor spectrum). The $p$ remains uncertain though various turbulence theories suggest a numerical value of $p=2$ (Saffman spectrum), $p=8/3$ (von K$\rm{\acute{a}}$rm$\rm{\acute{a}}$n spectrum) or $p=4$ (Batchelor spectrum) reviewed elsewhere \citep{pope2001turbulent}.  All theories agree that $p>1$ to ensure that as $k \rightarrow 0$, both $E_{ww}(k)\rightarrow 0$ and $(dE_{ww}(k)/dk)\sim k^{p-1}\rightarrow 0$. Not withstanding this uncertainty in $p$, its precise numerical value does not alter the scale-wise integrated outcome. A plausibility check on the assumed shape of $E_{ww}(k)$ is conducted using the integral constraint $\sigma_w^2=\int_{0}^{\infty} E_{ww}(k)dk$, yielding
\begin{eqnarray} \label{eq:sigmaus}  
\sigma_w^2
\approx \left[1.63-\frac{0.65p}{p+1}\left(\frac{l_o}{R}\right)-0.69\left(\frac{\eta}{l_o}\right)^{2/3}\right]u_*^2,
\end{eqnarray}
where a balance between production and dissipation of TKE yields $\epsilon(z_*) \approx u_*^2\Gamma(z_*)$ and $\Gamma(z_*)=u_*/l_o$. Equation \ref{eq:sigmaus} predicts a maximum $\sigma_w/u_*=\sqrt{1.63}=1.28$ sufficiently close to the reported $1.1-1.25$ range in near-neutral atmospheric flows, open channels, and pipes \citep{Raupach81,RaupachEA91,PoggiEA02,katul1996investigation,coscarella2021relation}.

Equation \ref{eq:fuw} can be further analyzed for the much-studied ISR and is shown to be consistent in both scaling law and similarity coefficients with accepted theories and experiments \citep{pope2001turbulent,saddoughi1994local}. For example, in the ISR, $\Phi(k)\ll1$, and the co-spectrum  reduces to 
\begin{equation}
F_{uw}(k)=\sqrt{C_o}\frac{(1-C_I)}{C_R}\Gamma (z_*) \left[\epsilon(z_*)\right] ^{1/3} k^{-7/3},     
\end{equation}
consistent with well-accepted co-spectral theories predicting $k^{-7/3}$ scaling \citep{pope2001turbulent,lumley1967similarity}. The emerging constants $\sqrt{C_o}{(1-C_I)}/{C_R}=0.18$ is also close to the accepted similarity constant reported in laboratory and field experiments as well as direct numerical simulations ($0.15$) discussed elsewhere \citep{wyngaard1972cospectral,saddoughi1994local,bos2004behavior,katul2013co}. These findings indirectly support setting $T_{uw}(k)=0$ for all $k$ as a first-order approximation in two ways: (i) its expected zero value in the ISR is needed to recover the $k^{-7/3}$ scaling, and (ii) $T_{uw}(k)$ must satisfy the integral constraint $\int_{0}^{\infty} T_{uw}(k)dk=0$ by definition.  In the case of $E_{kol}(k)$, the transfer of energy across scales shapes the energy cascade and is thus necessary for obtaining the $k^{-5/3}$ scaling in the ISR.  The inclusion of the transfer term in the energy cascade (indirectly specified by $E_{ww}(k)$) but not in the CSB may appear paradoxical.  This is not so as the role and significance of the transfer terms are quite different when analyzing scale-wise energy and scale-wise stress budgets \citep{bos2004behavior}.  Last, when $\Gamma(z)=0$, $F_{uw}(k)=0$ at all $k$.  Hence, a finite $\Gamma(z_*)$ is necessary to maintain a finite co-spectrum at all $k$ and $f_t>0$.

Returning to the determination of $f_t$, upon inserting equation \ref{eq:ewweq} into \ref{eq:fuw} yields the near-bed shear stress in terms of $f_t(=8\tau_*/\rho_f U_b^2)$ as
\begin{eqnarray} 
\label{eq:ff1}  
\frac{f_t U_b^2}{8}&
=& \frac{\xi(z_*)}{A_{\pi}}\left[
\int_{0}^{k_c} {{{k_c^{-p-2/3}k_o^{-5/3}k^{p}}}}dk +
\int_{k_c}^{k_o} {{{k_o^{-5/3}k^{-2/3}}}}dk \right. \nonumber\\
&+&
\left.\int_{k_o}^{k_e}
{{{k^{-7/3}}}}dk +B_{\pi}\int_{k_e}^{+\infty}
{{{k_e^{4/3}k^{-11/3}\exp(-\beta k \eta)}}}dk
\right],
\end{eqnarray}
\noindent where $A_{\pi}={C_R}/{\left[\sqrt{C_o}(1-C_I)\right]}\approx 5.58$, $B_{\pi}=\left({5^{4/3}}/{2}\right) C_R\sqrt{C_o} \approx 6.20$, and $\xi(z_*)=\Gamma(z_*)[\epsilon(z_*)]^{1/3}$. The four integrand functions are contributions to the turbulent stress arising from the two energetic, inertial, and dissipation ranges, respectively. 

\section{Results}
\subsection{Linking local and bulk variables}
The terms $\Gamma(z_*)$ and $\epsilon(z_*)$ needed in $\xi(z_*)$ are defined at $z_*$ and must be linked to bulk variables to complete the CSB model for $f_t$.  These are commonly estimated as \citep{bonetti2017manning,coscarella2021relation}
\begin{eqnarray}    
\label{eq:tauep}  
\Gamma(z_*)=\frac{u_*}{l_o}=c_t\frac{U_b-0}{l_o}, ~ \epsilon(z_*)=\frac{u_*^3}{l_o}=c_p^3 \frac{U_b^3}{R}, 
\end{eqnarray}
where $c_t(z_*)$ and $c_p(z_*)$ are unknown positive coefficients that link local to bulk variables given by $ c_t={u_*}/{U_b}$ and $c_p=({u_*}/{U_b})\left({R}/{l_o}\right)^{1/3}=c_t\left({R}/{l_o}\right)^{1/3}$. A bulk dissipation proportional to $U_b^3/R$ is compatible with upper limits set by prior variational analysis \citep{doering1994variational}.  Clearly, $c_t$ and $c_p$ cannot be individually constant and must vary with $f_t$ \citep{bonetti2017manning,coscarella2021relation}. The interest here is not in their individual variations but in their product. Increasing $l_o$ increases $c_p$ (more dissipation for the same $U_b$ or flow rate) but decreases $c_t$ because $U_b$ overestimates $U(z_*)$ thus making their product less sensitive to $l_o$ as shown elsewhere \citep{bonetti2017manning,coscarella2021relation}. 

For guessing a $c_t c_p$, several possibilities exist including the use of complete and incomplete similarity or covariate analysis \citep{barenblatt1995does}.  Another naive possibility is to assume $c_t c_p$ varies with the primary variable $R/l_o$ and proceed to select a maximum $c_t c_p$ at a given $R/l_o$.  By definition, $c_t c_p \sim ({u_*}/{U_b})^2 (R/l_o)^{1/3}$. With $s=R/l_o$, assuming $G_f(s)=({u_*}/{U_b})^2$ and maximizing $c_t c_p$ at a given $s$ leads to   
\begin{equation}
\label{eq:MEP}
\frac{d }{ds}(c_t c_p)= \left[\frac{G_f(s)}{3 s^{2/3}} + s^{1/3} \frac{d}{ds}G_f(s) \right]=0.
\end{equation}
The solution of equation \ref{eq:MEP} is $G_f(s)=b_o s^{-1/3}$, where $b_o$ is a constant independent of $s=R/l_o$. This argument is congruent with complete similarity theory in the limit of very large $R/l_o$ but cannot be correct for all $R/l_o$. Accepting momentarily a constant $c_t c_p$ at its maximal value, $f_t$ can be linked to $r/R$ and $Re$ at any finite $R/l_o$ using %
\begin{equation} 
f_t = \frac{30 (c_t c_p)}{A_{\pi}}\left[-Y(p) \left(\frac{k_c}{k_o}\right)^{2/3}+ \left(\frac{k_c}{k_o}\right)^{1/3}
-C_{\pi}\left(\frac{k_c}{k_e}\right)^{1/3}\frac{k_o}{k_e}\right],
\label{eq:fRIC}  
\end{equation}
where 
\begin{equation} 
Y(p)=\frac{12p+8}{15(p+1)}, ~C_{\pi}=\frac{1}{5}-{\frac {4\,\sqrt [3]{5}}{1875}{\beta}^{{\frac{8}{3}}}
\Gamma_{o,*} \left( -{\frac{8}{3}},{\frac {\beta}{5}} \right)},
\nonumber  
\end{equation}
\noindent $\Gamma_{o,*}(.)$ is the Gamma function and $C_{\pi}\approx 0.146$. For $p=2-4$, $Y(p)\approx0.72-0.75$ and variations in $p$ are hereafter ignored. The two extreme cases, Strickler and Blasius scaling are now evaluated. In a rough pipe where $r/L_v\gg1$, $DS(k)$ can be ignored and $k_e=k_o (r/\eta) \rightarrow \infty$ allowing the IR to extend to $k_e\rightarrow \infty$. In the limit of $r/R\ll1$, the leading order term in equation \ref{eq:fRIC} is
\begin{equation} 
\label{eq:ffrg1}    
f_t \approx \frac{30}{A_{\pi}}c_pc_t \left(\frac{r}{R}\right)^{1/3}.
\end{equation}
Hence, the Stickler scaling requires: i) a constant $c_p c_t$ ($=A_sA_{\pi}/30 \approx 0.026$ to recover the Nikuradse data), ii) $r/L_v \gg 1$, and iii) $r/R\ll 1$. Likewise, when $r/L_v \ll 1$ so that $l_o \approx L_v$, the inertial subrange commences at $k_e$ and rapidly terminates into a dissipation range since $k_r/k_e \ll 1$. The viscous cutoff effects become important when $k>k_e$ revising equation \ref{eq:fRIC} to 
\begin{equation} \label{eq:ffsm}
{f_t}
\approx \frac{30D_{\pi}}{A_{\pi}}c_tc_p \left(\frac{\eta}{R}\right)^{1/3} =
\frac{30D_{\pi}}{2^{-1/4}A_{\pi}}c_t c_p^{3/4}Re^{-1/4},
\end{equation}
where $D_{\pi}=\sqrt[3]{5}-C_{\pi}/5\approx1.68$, $(R/\eta)^{1/3}=(c_p Re/2)^{1/4}$. The Blasius scaling requires $c_t c_p^{3/4}$ ($=2^{-1/4}A_bA_{\pi}/(30D_{\pi}) \approx0.0294$ for Nikuradse data) not to vary with $Re$ (or equivalently $c_t c_p$ not to vary with $R/l_o$ when $l_o=L_v$ as before).  Equation \ref{eq:fRIC} allows separating the effects of turbulent exchanges of momentum at $z_*$ from relations between local (at $z_*$) and bulk variables (encoded in $c_t c_p$) when evaluating $f_t$ or NG06. 

\subsection{Solution of the implicit function in NG06}
The study objective, which is to derive the $g_o(\chi)$ in NG06 for the Nikuradse data ($r/R\ll1$) and regular roughness, can now be addressed.  The $g_o(\chi)$ can be made explicit when rearranging equation \ref{eq:fRIC} to yield, 
\begin{eqnarray} 
\label{eq:ffgn}  
g_o(\chi)=(c_t c_p^{3/4})\frac{30}{2^{-1/4}A_{\pi}}\left\{
\left[\left(\frac{c_p}{2}\right)^{3/4}\chi+5\right]^{1/3}-C_{\pi}\left[\left(\frac{c_p}{2}\right)^{3/4}\chi+5\right]^{-1}
\right\},
\end{eqnarray}
where $\chi=Re^{3/4}\left({r}/{R}\right)$ derived from $\chi=(2/c_p)^{3/4}(r/\eta)$. Now equation \ref{eq:ffgn} explains why the Nikuradse data imperfectly collapses along a unique curve when plotting $f_t Re^{1/4}$ versus $Re^{3/4}\left({r}/{R}\right)$ under the restrictive assumption of constant $c_tc_p$. Thus, the main novelty here is to show that the $g_o(\chi)$ in NG06 can be linked to an approximated Navier-Stokes equation (i.e. the CSB model) provided $c_t c_p$ is constant at maximal value, which is the sought result. The solution of equation \ref{eq:ffgn} is also presented in Figure \ref{fig:univ1} where $c_p=2(D_{\pi}A_{s}/A_{b})^4\approx0.6$ and can be directly derived when combing the Strickler and Blasius scaling laws. Moreover, a sensitivity analysis was conducted by setting $c_p=0.3, 1.2$ to find the best fit between equation \ref{eq:ffgn} and the Nikuradse data set. Figure \ref{fig:univ1} shows that setting $c_p$ as constant (accepting Blasius and Strickler scaling laws simultaneously) can indeed replicate the NG06 result to a leading order, but further investigations are needed when $\chi<20$. 

\section{Discussion}
\subsection{Extension to micro-scale and  large-scale roughness}

Moving beyond the widely-used Nikuradse data range for $r/R< 0.1$ and for regular roughness elements, the following discussion are presented to assess the plausibility of extending the implicit function $g_o(\chi)$ to two extreme cases as shown in Figure \ref{fig:univ}: i) a hydrodynamically smooth regime or the micro-scale roughness $r/R\in[10^{-6}, 10^{-5}]$ from Hi-Reff \citep{furuichi2015friction} super-pipe experiments (the Oregon and Princeton \citep{swanson2002pipe, MckeonEA04} are assumed smooth though no $r/R$ measurements were reported), and (ii) a large-scale roughness regime $r/R\in[0.1, 0.2]$ from pipes roughened with single layers of sand \citep{huang2013experimental}. The reported friction factor data \citep{huang2013experimental} (runs $R4$ and $R5$) in the original Table 1 were employed. These two runs can still be approximated as regular roughness with $r/R$ not too large so that the prior conditions imposed on $z_*$ for the use of the CSB can still be enforced.

\subsection{Estimation of $c_tc_p$}
\begin{figure} 
\centerline{\includegraphics[angle=0,width=0.99\linewidth]{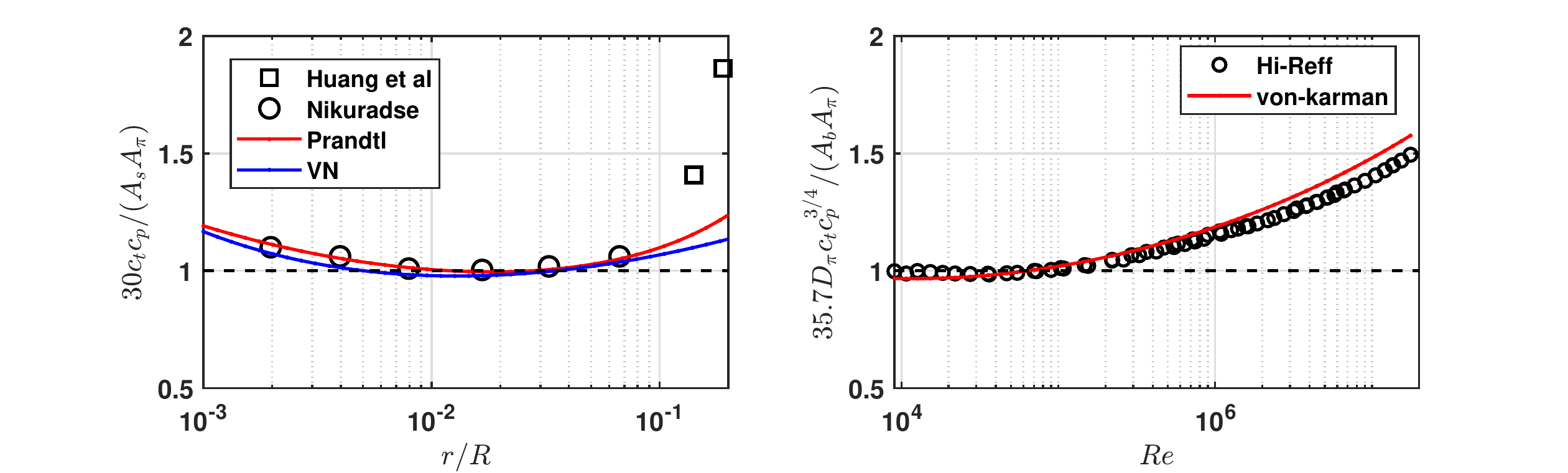}}
\caption{Exploring the similarity coefficient of Strickler and Blasius scalings. The data sets are normalized by $A_s$ and $A_b$, the similarity coefficients for the Strickler and Blasius formulae, respectively where the purple line highlights the unit line. Other than the experimental data, the blue lines represents the predictions from Prandtl and von K$\rm{\acute{a}}$rm$\rm{\acute{a}}$n equations (or the Colebrook-White formula when added), and the black dashed line features the prediction from virtual Nikuradse curves. }
\label{fig:cns}
\end{figure}

To extend the proposed model and without invoking further \textit{ad hoc} assumptions on the local flow structure, a 'naive' but direct approach is to revise the constant $c_t$, $c_p$ assumption that seems only applicable to the Nikuradse range \cite{gioia2001scaling,coscarella2021relation}.  When inferring $U_b$, the log-law is assumed to populate $U^+$ over extensive portions of the pipe area at intermediate to high $Re$.  The log-law overestimates $U^+$ in the buffer region (for smooth pipes) or the roughness sublayer (in rough pipes) but underestimates $U^+$ in the wake-region \citep{katul2014cospectral}.  Thus, its area-integrated form from $l_o$ to $R$ may be less sensitive to such deviations and provides a leading order guess as to whether $c_t c_p$ is constant or variable. With this idealized $U^+$ representation, it follows that $c_t^{-1}\approx {U_b}/{u_*} = ({1}/{\kappa})\ln\left({R}/{l_o}\right)+B_o$, where $B_o$ is an integration constant of order unity and $\kappa$ is the von K\'{a}rm\'{a}n constant \citep{pope2001turbulent}. For this $c_t$ and $c_p=c_t (R/l_o)^{1/3}$, their product can now be estimated as 
\begin{equation} \label{eq:Ub_log2}  
c_t c_p\approx \left(\frac{R}{l_o}\right)^{1/3} \left[\frac{1}{\kappa}\ln\left(\frac{R}{l_o}\right)+B_o\right]^{-2},
\end{equation}
Increasing $R/l_o$ increases both numerator and denominator thereby making their ratio less sensitive to $R/l_o$ as expected.  However, a near constant $c_t c_p$ emerges when noting that for large but finite $R/l_o$, $\ln(R/l_o) \approx A_n (R/l_o)^{n_o}$ with $n_o=1/6$ for the range covered by $R/l_o$ in many experiments \citep{katul2002mixing}. To elaborate, the limiting case for large $R/l_o$ is considered and this case leads to 
\begin{equation} \label{eq:no_high}  
n_o \approx \lim_{R/l_o\rightarrow\infty} \frac{\log[\log(R/l_o)]}{\log(A_n)+\log(R/l_o)}\approx \frac{1}{\log(R/l_o)}
\end{equation}
when applying L'H$\rm{\hat{o}}$pital's rule. The $n_o = 1/\log(R/l_o)$ appears independent of $A_n$ but weakly depends on $R/l_o$ as shown from a similar argument using asymptotic covariance analysis \citep{barenblatt1995does}.  In general, ${n_o}\rightarrow 1/\log(R/l_o)$ for very large $R/l_o$ and cannot be a constant. To explore the plausibility of setting $c_tc_p$ a constant beyond the Nikuradse  experiments, other predictions from the virtual Nikuradse \citep{yang2009virtual} equation (VN), the Moody diagram summarized by the approximate von K$\rm{\acute{a}}$rm$\rm{\acute{a}}$n equation \citep{Prandtl}), and the aforementioned micro-scale and large-scale roughness data are employed and discussed in Figure \ref{fig:cns}.

Figure \ref{fig:cns} shows that $c_t c_p$ does not vary appreciably with $r/R$ for small-scale roughness ($r/R<0.1$) consistent with the range of applicability \citep{bonetti2017manning,coscarella2021relation}. However, as $r/R$ increases to 0.2, $c_t c_p$ increases leading to a break-down in the Strickler scaling.  This breakdown originates from estimates of $\xi(z_*)$ when using bulk variables and not in the particulars of momentum exchange by turbulent eddies at $z_*$ represented by the CSB model. Likewise, for the Blasius scaling the $c_t c_p^{3/4}$ (or $c_t c_p$ independent of $R/L_v$) remains flat for a restricted range of $Re \in[10^{4}, 10^5]$ but increases significantly with increasing $Re$. 

\begin{figure}
\centerline{\includegraphics[angle=0,width=0.7\linewidth]{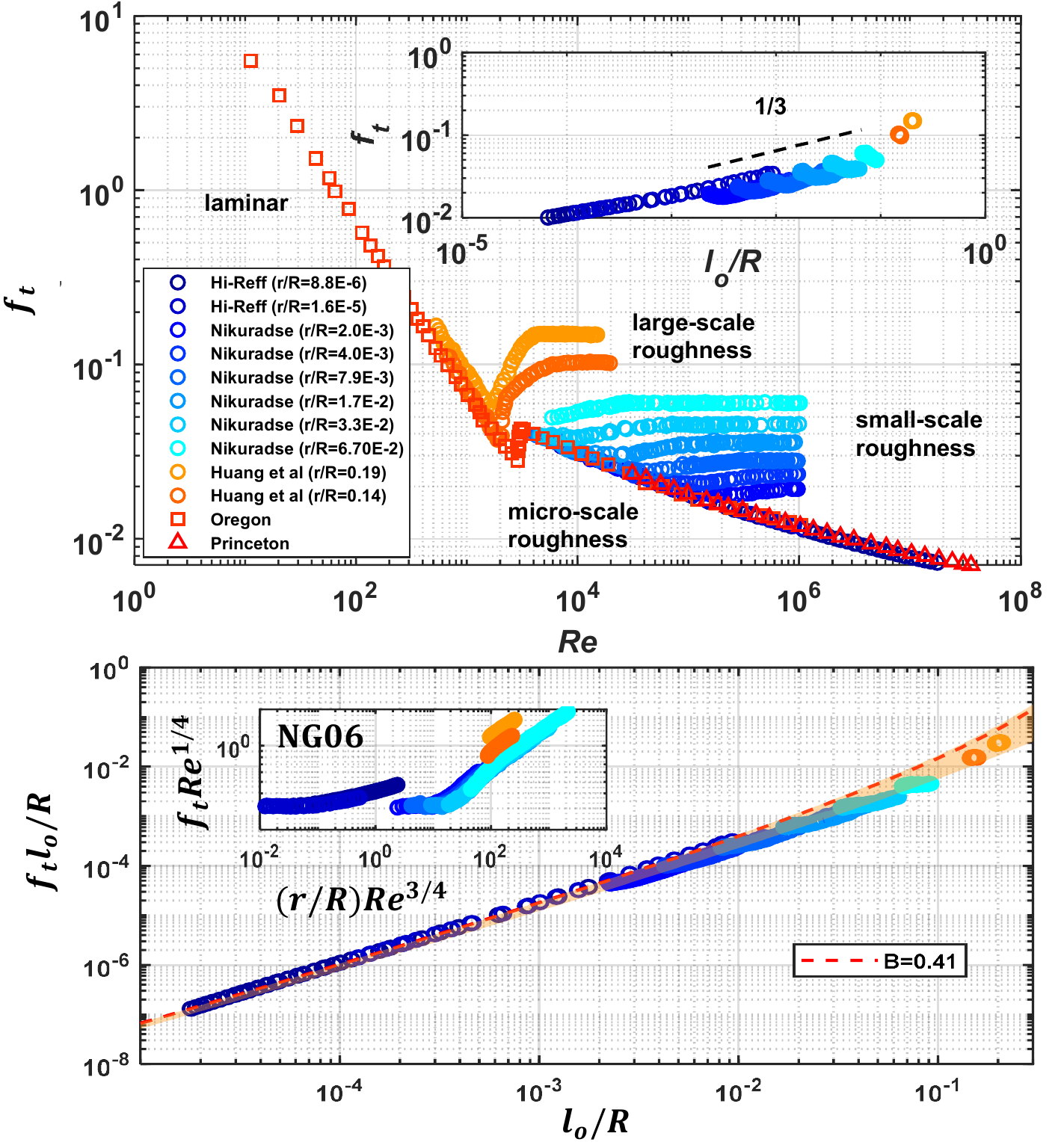}}
\caption{Representation $f_t$ across regular roughness elements. The top panel shows the expanded Nikuradse diagram using wider experimental conditions, including the large-scale roughness \citep{huang2013experimental}, intermediate/small-scale roughness \citep{Nikuradse}, and smooth-wall \citep{swanson2002pipe, MckeonEA04, furuichi2015friction}. The inset is a representation of all $f_t$ data with $l_o/R$ as a general scale using the turbulence data (excluding non-turbulent regimes). The bottom panel shows the data collapse of $f_t (l_o/R)$ versus $l_o/R$ where the orange band features CSB prediction with $B_o\in[0.04, 4]$. The inset is the NG06 curve for the same data.}
\label{fig:univ}
\end{figure}

With modeled $c_tc_p$ provided in equation \ref{eq:Ub_log2}, a representation of extended Nikuradese diagram with additional micro-scale and large-scale roughness data is shown in Figure \ref{fig:univ}.
Figure \ref{fig:univ} shows that all these $f_t$ data reasonably collapse along a one-dimensional curve (predicted from CSB) when plotted versus $l_o/R$.  This finding indicates that $l_o/R$ is a characteristic length scale that describes $f_t$ in all regimes as alluded to in earlier studies \citep{bonetti2017manning,coscarella2021relation}. Similar to the data collapse strategy that NG06 employed, an apparent curve can also be derived when plotting $f_t \cdot (l_o/R)$ versus $l_o/R$, where $f_t \cdot (l_o/R)$ can be understood physically as a 'roughness friction factor' noting that  $f_t (l_o/R)=2g l_o S_b /U_b^2$. The improved data collapse from the $f_t \cdot (l_o/R)$ representation is partly connected to self-correlation because the abscissa and ordinate now share the same variable $(l_o/R)$ that span several orders of magnitude.  Likewise, the NG06 representation also suffers from similar self-correlation through $Re$, which varies over several orders of magnitude as well.  This finding confirms the applicability of the proposed CSB model at the two extremes of $(l_o/R)$ albeit models for $c_tc_p$ are required as deviations from a constant product value are expected.  These deviations are connected to how bulk variables relate to local mean velocity gradient and TKE dissipation rate at $z_*$ instead of how eddies transport momentum to pipe walls at $z_*$.

\section{Conclusion}
An explicit solution for the NG06 conveyance equation for friction factor, originally conjectured from analogies to critical phenomenon, was derived from a CSB model. The CSB model employs standard turbulent theories and a commonly accepted wall-normal velocity spectrum. The model closes the pressure-velocity de-correlation term using a linear Rotta scheme based on linear return-to-isotropy with adjustments due to isotropization of the production term. Moving beyond and above the CSB model, the extension of the CSB prediction is also discussed in terms of micro-scale roughness and large-scale roughness experiments that were not covered by the original Nikuradse range. The analysis shows that all the turbulent friction factor data collected so far can be approximately collapsed onto a single curve. However, the work here shows that much of the uncertainty originates from how local to bulk variables are related instead of the mechanics of momentum exchange with the pipe walls. 

\textbf{Acknowledgements}: Support from the U.S. National Science Foundation (NSF-AGS-1644382, NSF-AGS-2028633, and NSF-IOS-1754893) is acknowledged.

\bibliographystyle{jfm}
\bibliography{2_CSB_ft}
\end{document}